\documentclass[aps,prc,twocolumn,showpacs,preprintnumbers,amsmath,amssymb,epsf,superscriptaddress,groupedaddress,nofootinbib,tightenlines]{revtex4}

\usepackage{graphicx}
\usepackage{bm}
\usepackage{amsmath}
\usepackage{amssymb}
\usepackage{latexsym}

\renewcommand{\vec}{\boldsymbol}
\newcommand{\mc}{M_{\mathrm{c}}}
\newcommand{\mn}{M_{\mathrm{n}}}
\newcommand{\mnc}{M_{\mathrm{nc}}} 
\newcommand{\mr}{M_{\mathrm{R}}}
\newcommand{\gma}{\gamma}
\newcommand{\gmat}{\tilde{\gamma}}
\newcommand{\pc}{\vec{p}_{c}}

\newcommand{\lis}{\ensuremath{{}^{7}\mathrm{Li}}} 
\newcommand{\lie}{\ensuremath{{}^{8}\mathrm{Li}}} 
\renewcommand{\S}[2]{{}^{#1}S_{#2}}
\renewcommand{\P}[2]{{}^{#1}P_{#2}}
\newcommand{\gone}{g_{(\S{3}{1})}}
\newcommand{\gtwo}{g_{(\S{5}{2})}}
\newcommand{\gthree}{g_{(\S{3}{1}^{*})}}
\newcommand{\aone}{a_{(\S{3}{1})}}
\newcommand{\atwo}{a_{(\S{5}{2})}}
\newcommand{\hone}{h_{(\P{3}{2})}}
\newcommand{\htwo}{h_{(\P{5}{2})}}
\newcommand{\hpt}{h_{t}}
\newcommand{\hthree}{h_{(\P{3}{2}^{*})}}
\newcommand{\honet}{\tilde{h}_{(\P{3}{1})}}
\newcommand{\htwot}{\tilde{h}_{(\P{5}{1})}}
\newcommand{\hthreet}{\tilde{h}_{(\P{1}{1}^{*})}}
\newcommand{\hfourt}{\tilde{h}_{(\P{3}{1}^{*})}}
\newcommand{\htott}{\tilde{h}_{t}}
\newcommand{\htotst}{\tilde{h}_{t^{*}}}

\newcommand{\V}[1]{\vec{V}_{#1}}
\newcommand{\fdu}[2]{{#1}^{\dagger #2}}

\newcommand{\fu}[2]{{#1}^{#2}}
\newcommand{\fd}[2]{{#1}_{#2}}
\newcommand{\T}[2]{T_{#1}^{\, #2}}
\newcommand{\e}{\vec{\epsilon}}
\newcommand{\es}{\e^{*}}
\begin{document}

\title{Marrying {\it ab initio} calculations and Halo-EFT: the case of ${}^7{\rm Li} + n \rightarrow {}^8{\rm Li} + \gamma$}

\author{Xilin Zhang} 
\author{Kenneth M.~Nollett} 
\author{D.~R.~Phillips} 
\affiliation{Institute of Nuclear and Particle Physics and Department of
Physics and Astronomy, Ohio University, Athens, OH\ \ 45701, USA}

\date{Aug, 2013}

\begin{abstract}
We report a leading-order calculation of radiative $\lis$
neutron captures to both the ground and first excited
state of $\lie$ in the framework of a low-energy effective field
theory (Halo-EFT). Each of the possible final states is treated as a
shallow bound state composed of both $n+\lis$ and $n+\lis^{*}$ (core
excitation) configurations.
The {\it ab initio} variational Monte Carlo method is used to compute
 the asymptotic normalization coefficients of these bound states, which are then used to fix couplings in our EFT. We calculate the total 
and partial cross sections
in the radiative capture process using this calibrated EFT.
Fair agreement with measured total cross sections is achieved and excellent 
agreement with the measured branching ratio between the two final states is found.
In contrast,a previous Halo-EFT calculation [G.~Rupak and R.~Higa, Phys. Rev. Lett {\bf 106}, 222501 (2011)] assumes that the  $n$-$\lis$ couplings in different spin channels are equal, fits the $P$-wave ``effective-range" parameter
to the threshold cross section 
for ${}^7{\rm Li} + n \rightarrow {}^8{\rm Li} + \gamma$, and assumes the core excitation is at high enough energy scale that it can be integrated out.  

\end{abstract}
\maketitle

\section{Introduction} \label{sec:intro}

\subsection{Motivation}

The radiative capture of neutrons on \lis~[denoted $\lis(n,\ \gma)\lie$] 
has been studied both experimentally
\cite{Blackmon,Barker,Lynn,Nagai,knox87,Heil,Imhof} and theoretically
\cite{Davids:2003aw,Tombrello,Aurdal,Jennings,Huang,Navratil:2010jn,Rupakprl,Fernando:2011ts,Esbensen}
for many years. The reaction could be important for the $r$-process in 
neutrino-driven winds,
as part of a path around the $A=5$ and $A=8$ stability gaps \cite{Sasaqui}, although 
most contemporary interest focuses on the use of $^7\mathrm{Li}(n,\gamma)^8\mathrm{Li}$ to 
constrain models of the isospin mirror process 
$^7\mathrm{Be}(p,\gamma)^8\mathrm{B}$~\footnote{The reaction 
$^7\mathrm{Li}(n,\gamma)^8\mathrm{Li}$
is also important in inhomogeneous models of big-bang nucleosynthesis
\cite{Heil,Malaney}, but these are now disfavored because they
overproduce several light nuclides and because the laboratory evidence
is now against a first-order QCD phase transition in the early
universe.}.
The flux of solar neutrinos, especially at the energies probed by
water Cherenkov and chlorine experiments, is sensitive to the rate of
the proton-capture reaction~\cite{Adelberger:2010qa}.
It is at energies typical of the solar interior ($\sim
20$ keV) that the cross section must be known,
and the relevant reaction mechanism there is nonresonant direct
capture. Since the cross section is unmeasurably small in the 10s of keV
regime models
are necessary to combine experimental cross sections measured at
higher energies and extrapolate them to energies found inside
the Sun. But even
measurement of the $^7\mathrm{Be}(p,\gamma)^8\mathrm{B}$
rate at energies of 100s of keV is difficult because of its small cross sections
and radioactivity of the $^7$Be target; large uncertainties have
lingered for a long time \cite{Adelberger:2010qa}.  

Many theoretical models of the $^7\mathrm{Be}(p,\gamma)^8\mathrm{B}$
reaction have been put forward.  (For a thorough discussion and an
exhaustive list of references, see Ref.~\cite{Adelberger:2010qa}.)
Like the experimental effort that they support, the models face severe
practical limitations.  Simplest in conception are ``two-body''
potential models, in which the $^7$Be nucleus and captured proton are
treated as fundamental particles interacting through a Woods-Saxon or
similar potential.  These models suffer from a lack of experimental
constraints, since low-lying states may
not be relevant to the channels probed in the reaction.  Several
attempts have been made to constrain models by studying the \lie\ 
system and applying isospin symmetry, using both Woods-Saxon and
$R$-matrix models
\cite{Tombrello,Aurdal,Barker,Esbensen,Timofeyuk,Trache}.  These met
with only partial success, generally failing to reproduce absolute
cross sections of $^7\mathrm{Be}(p,\gamma)^8\mathrm{B}$ and
$\lis(n,\gamma)\lie$ simultaneously.  The source of this difficulty
remains unclear~\cite{Esbensen}.

Another path has been to compute cross sections in models that contain
nucleon degrees of freedom in the target, linking the cross section to
the nucleon-level physics.  Until recently, computational limits
forced rather severe simplifications on such ``microscopic'' models.
They do not typically predict the absolute cross section accurately
but seem to give a better description of its energy dependence than do potential models~\cite{Johnson:1992,Descouvemont:1994,Descouvemont:2004}.  
Thus the overall scale of these results was often adjusted by fitting to data. (See also Refs.~\cite{Bennaceur,Halderson:2006ru}
for similar results using methods related to the ``traditional'' shell model.)

However, recent advances in computing power and descriptions of the
nucleon-nucleon interaction have made possible \textit{ab initio}
models based on detailed descriptions of the underlying
dynamics.  Accurate \textit{ab initio} calculations of $A=8$ bound and
resonant states have been available for more than a decade
\cite{Wiringa:2000gb,Navratil:2003}, and completely \textit{ab initio} calculations of a few
reaction processes \cite{Nollett:2007,Quaglioni:2008,Navratil:2010jn} are beginning to emerge.
In particular, Ref.~\cite{Navratil:2011sa} presented a well-converged computation of
$^7\mathrm{Be}(p,\gamma)^8\mathrm{B}$ with what should be a suitably
accurate description of the nucleon-nucleon interaction. Thus, in contrast to older calculations with nucleonic 
degrees of freedom, these studies provide information on
cross sections that is not in any way fitted to the capture data.
\textit{Ab initio} and experimental cross sections can now be compared
directly to one another, with the promise of more accurate input to, e.g., solar models, 
in the future.

Nevertheless, the \textit{ab initio} models' margin of advantage over potential
models remains relatively small: the energy dependences of non-resonant cross
sections are dominated by barrier penetrabilities and phase-space-type
considerations. These features are captured in any
reasonable model.  There should consequently be considerable utility in finding
ways to extract irreducibly nucleon-level information from \textit{ab
  initio} methods and use it in less computationally-intensive models
that describe ``non-microscopic'' effects directly and transparently.
The initial application of \textit{ab initio} methods to radiative
captures indeed followed this path, combining radial overlap
functions (or spectroscopic factors) from nuclear structure models
 with simpler models of nucleus-nucleus scattering~\cite{Barker,Nollett:2001a,Nollett:2001b,Navratil:2006tt}.

\subsection{Why effective field theory?}

But a serious weakness of such a procedure is that it has 
uncontrolled errors.  For example, in a Woods-Saxon model
with one open channel there are (minimally) four parameters to be
fitted: central and spin-orbit potential strengths, and two geometric
parameters.  The parameters are fitted to whatever information may be
available, and then ambiguities in the parameters translate to
uncertainties in the computed cross sections.  Whether the Woods-Saxon
model provides the best possible constraint on cross sections given
input data is unclear, and there is no systematic approach to
determine what is allowed.  Past efforts to assess 
the theoretical uncertainties associated with potential-model
approaches in the $A=8$ system may be found in Refs.~\cite{Jennings,Davids:2003aw}.

Effective field theory (EFT)
provides an alternative approach.  Rather than an assumed form for
the two-body potential, it starts with a Lagrangian for the motion and
interaction of the colliding particles.  The Lagrangian is expanded in
powers of a small momentum scale characterizing the problem, including
all operators consistent with the underlying symmetries up to
some specified power of the momentum scale. The presence of low-energy bound states is built into the EFT by assumptions
about the size of the coefficients that govern the interaction, which are 
 fitted to match
input data. Observables that are not amongst the input data can
then be predicted (or postdicted).  This provides a systematic sorting of physical effects
by their importance and builds in a minimum set of assumptions. 
For reviews
describing the application of such EFTs to nuclear physics see Refs.~\cite{Beane:2000fx,Bedaque:2002mn,Epelbaum:2008ga,HP10}

In this case the degrees of freedom in the EFT will be the halo nucleus' core and the valence neutrons,
supplemented by degrees of freedom associated with core excited states. The EFT expansion in
halo systems is an expansion in $\gamma/\Lambda$, where $\gamma$ is the binding momentum associated
with the halo bound (or scattering) state, and $\Lambda$ is a high-momentum scale related to the range of
the core-neutron interaction. 
The resulting ``Halo-EFT" approach thus contains similar physics to a potential model,
but includes fewer tacit assumptions about the system and a better
accounting of uncertainties.  In particular, the size of the error due to omitted
higher-order operators can be estimated quantitatively.  The main
requirement for a successful Halo-EFT treatment is that the characteristic momentum of the halo
physics ($\gamma$) be well below the scale at which details of the neutron-core
interaction become important. 
Similar assumptions inhere to potential models, and the role
of ``simple'' physics like barrier penetrabilities and phase space
should be very similar in the two approaches. Halo-EFT has its genesis in
the ``short-range EFT" (sometimes called ``pionless EFT") originally developed in 
Refs.~\cite{vK99,Ka98A,Ka98B,Ge98,Bi99}, with that theory extended to $P$-wave
interactions in Refs.~\cite{Be02,Bd03} and the results applied to various
halo systems including ${}^5$He~\cite{Be02,Bd03}, ${}^8$Li~\cite{Rupakprl,Fernando:2011ts}, ${}^{11}$Be~\cite{Hammer:2011ye},
${}^{15}$C~\cite{Rupak:2012}, ${}^{17}$F~\cite{Ryberg:2013} and ${}^{19}$C~\cite{Acharya:2013}, as well as to $\alpha$-$\alpha$
interactions~\cite{Higa:2008} and a number of two-neutron halos~\cite{Canham:2008,Canham:2009,
Rotureau:2013,Acharya:2013B,Hagen:2013,Hagen:2013B}. 

None of these calculations, however, used input from \textit{ab initio} computations. 
  In fact, practitioners of \textit{ab initio} models have
been computing results that can be used quite directly to fix
Halo-EFT parameters: nuclear asymptotic normalization coefficients
(ANCs) \cite{Timofeyuk:2010}, which are quite often used
as inputs to potential models of
radiative capture~\cite{Xu,Huang}.
Relative to \textit{ab initio} calculations, both potential
models and Halo-EFT treat the simple physics simply.  Halo-EFT should be the 
better framework for doing this, because it provides 
physically motivated 
calculations consistent
with \textit{ab initio} theory which have a  known level of accuracy.

\subsection{Degrees of freedom and energy scales}

In this report, we describe an application of Halo-EFT to the $A=8$
capture process $\lis(n,\gma)\lie$. The degrees of freedom in our 
calculation are the neutron together with the ground and first-excited states of 
the $\lis$ core.  Halo-EFT has
previously been applied to this system
\cite{Rupakprl,Fernando:2011ts}, but, in contrast to those works we use ANCs from  \textit{ab initio}
calculations to fix most EFT parameters. 

In our work the high-energy scale is associated with the breakup energy of $\lis\rightarrow
t+{}^{4}\mathrm{He}$, $2.5$ MeV,
which translates to a high-momentum scale $\Lambda\sim 90$ MeV.  Starting from the binding energies
of $\lie$ and $\lie^{*}$ with respect to the $\lis$-$n$ threshold, $2.03$ and $1.05$ MeV respectively, we then 
infer that $\lie$ ($\lie^*$) is a bound state with typical momentum $\gma=57.8$ MeV ($\tilde{\gamma}=41.6$ MeV). (Various low energy/momentum scales that will be used later are collected in Table~\ref{tab:dynamicenergyscales}.)
These scales are then to be considered small with respect to $\Lambda$, which yields a nominal expansion parameter
$\gamma/\Lambda \sim 0.5$. However, the result we find for the $P$-wave effective range in $\lis$-$n$ scattering, $r_1 \approx -1.4$ fm$^{-1}$ 
suggests 
a higher $\Lambda$ and hence a more convergent
expansion. Since the theory is designed to work for bound-state energies of 1--2 MeV 
it should also describe $\lis$-neutron scattering provided the neutron energy is kept 
in this range. However, at approximately 
$0.22$ MeV above the 
 $n$-$\lis$ 
threshold, there is a $3^{+}$ resonance
unrelated to the threshold capture. This
resonance dominates the total cross section in a narrow window around 0.22 MeV. Since our goal is an understanding of
threshold capture (in particular with a view to computing radiative proton capture on ${}^7$Be) here  
we focus on 
the cross section below this energy and do not introduce this
resonance in our EFT. The 
  $3^{+}$ resonance could be added as an additional dynamical degree of freedom
  in the EFT, as was done in Ref.~\cite{Fernando:2011ts}.
  
  \begin{table}
  \begin{tabular}{|c|c|c|}
 \hline
 Momentum scale & Definition & Value\\
 \hline
 $\gamma$ & $\sqrt{2 M_R B_{\lie}}$ & 57.8 MeV\\
 
 $\gamma^*$ & $\sqrt{2 M_R(B_{\lie} + E^*)}$ & 65.1 MeV\\
 
 $\gamma_\Delta$ & $\sqrt{2 M_R E^*}$ & 30.0 MeV\\
 
 $\tilde{\gamma}$ & $\sqrt{2 M_R B_{\lie^*}}$ & 41.6 MeV\\
 
 $\tilde{\gamma}^*$ & $\sqrt{2 M_R (B_{\lie^*} + E^*)}$ & 51.3 MeV\\
 \hline
\end{tabular}
\caption{Momentum scales associated with bound states considered in our calculation. Note that the star ($^*$) denotes a binding energy or momentum computed relative to the $n$-$\lis^*$ threshold, which sits $E^*$ above the $n$-$\lis$ threshold, while quantities relevant to the $\lie$ excited state are indicated by a tilde ($\tilde{~}$). Note also that the definitions of, e.g. $\gamma$, $\gamma^*$, and $\gamma_\Delta$, result in an obvious Pythagorean equality. See the text below Eqs.~(\ref{eqn:selfEpi}) and~(\ref{eqn:Cs}) for a detailed discussion of these quantities.} \label{tab:dynamicenergyscales}
\end{table}
 
 \subsection{Differences from previous calculations}
There are three key differences between this study and the earlier Halo-EFT studies of this system: Refs.~\cite{Rupakprl,Fernando:2011ts}. 
First, we infer parameters of our leading-order (LO) EFT using binding
energies and $S$-wave scattering lengths taken from experiment and
\textit{ab initio} ANCs calculated by the variational 
Monte Carlo
(VMC) method \cite{Nollett:2011qf} with realistic two- and
three-nucleon interactions.
This explains the title of this article.  The
additional input information in our approach lets us predict ratios of 
partial cross sections as dynamical
quantities related to the couplings in the EFT Lagrangian.  In
contrast, the couplings of spin channels in the final state were
assumed
for simplicity
 to be equal in Ref.~\cite{Rupakprl,Fernando:2011ts} so that 
the branching
ratios there carry no link to the actual short-range physics at work in the $\lis$-neutron system. 

The second difference is the treatment of core excitation in the EFT.
The first excited state of \lis, with quantum numbers $J^\pi=1/2^-$ and denoted
here as $\lis^{*}$, is included as a dynamical degree of freedom in our
EFT. We see this as mandatory because its excitation energy  of $0.478$ MeV is small compared with
the neutron separation energy of $\lie$.  
As a result, low-energy $n$-$\lis$ scattering has an inelastic channel $n+\lis\rightarrow n+\lis^{*}$, and this reaction has been observed experimentally (e.g., see Ref.~\cite{knox87} and references therein). The importance of core excitation can also be seen by comparing
computed ANCs (see Table~\ref{tab:QMCANCs}) of $\lie$ for separation into
$n+\lis$ and into $n+\lis^{*}$: they are of similar size.  
We will show below that including core excitation 
yields a different conceptual picture of how the effective-range expansion (ERE) arises in the elastic channel.

Finally, we also provide predictions for the radiative capture of neutrons on ${}^7$Li into the first  excited state of ${}^8$Li, which is even shallower than the ground state, and so should be an even better candidate for the application of EFT.
This $1^+$ state was not considered in Ref.~\cite{Rupakprl}, and the branching ratio was not predicted but instead used as input
in the computation of Ref.~\cite{Fernando:2011ts}. We also include core excitation of $\lis$, i.e. effects of $\lis^* (1/2^-)$,
 when studying capture into $\lie^*$, just as we do for capture to the ground state. In principle similar calculations (not presented here) can be done to obtain LO Halo EFT predictions for the Coulomb dissociation of $\lie$ and $\lie^{*}$ to $\lis$ or $\lis^{*}$.  
  
The rest of the paper is organized as follows: Sec.~\ref{sec:example} uses a simple example to explain the idea of combining {\it ab initio} calculations (VMC) and the Halo-EFT to study two-body $P$-wave scattering and shallow bound states. In Sec.~\ref{sec:Li8bs}, we apply this idea to study $n$-$\lis$ ($\lis^{*})$ scatterings and $\lie\ (\lie^{*})$, with extra complexities including channel mixing and core excitation. In Sec.~\ref{sec:capture} we study the radiative captures at LO, discuss higher-order contributions, and compare our cross section results with available data. We summarize our study in Sec.~\ref{sec:sum}.   
\section{An example of a shallow $P$-wave bound state} \label{sec:example}
In this section we use a simple example to lay out the general idea of
studying a shallow $P$-wave bound state (``dimer") in Halo-EFT
\cite{Ka98A,Hammer:2011ye}. We only consider spinless particles; the generalization to degrees of freedom with spin will be performed in the next section.
The result will be a 
next-to-leading-order (NLO) calculation for which we can use ANCs computed via {\it ab initio} methods as input.  The EFT Lagrangian is
\begin{eqnarray}
\mathcal{L}_{0}&=&\fdu{n}{} \left(i\partial_{t}+\frac{\bigtriangledown^{2}}{2\mn} \right) \fd{n}{} + \fdu{c}{}\left(i\partial_{t}+\frac{\bigtriangledown^{2}}{2\mc} \right) \fd{c}{} \notag \\
&&{}+\fdu{\pi}{i}\left(i\partial_{t}+\frac{\bigtriangledown^{2}}{2\mnc}+ \Delta \right) \fd{\pi}{i} \notag \\
\mathcal{L}_{P}&=&h \fdu{\pi}{i}\fd{n}{} i\left(\V{n}-\V{c}\right)_{i} \fd{c}{}+\mathrm{C.C.} \ . \label{eqn:simpleL}
\end{eqnarray} 
Here the dimer fields, $\fdu{\pi}{i} \ (i=\pm1,0)$, correspond to the $P$-wave bound states, while $\fd{c}{}$ and $\fd{n}{}$ are the fields of the core and
the particle which bind to form the dimer. The masses in $\mathcal{L}_{0}$ are $\mn$ (neutron), $\mc$ (core), and $\mnc \equiv \mn + \mc$. 
$\Delta$ is the bare dimer binding energy, $h$ is the dimer-$c$-$n$ coupling, and $\V{c}, \ \V{n}$ are the core and neutron velocities~\footnote{Note the use of velocity couplings here, in contrast to the momentum couplings of Refs.~\cite{Be02,Bd03,Hammer:2011ye}.}

Now we calculate the self-energy of the dimer, which will be used to dress 
the dimer propagator, as shown in Fig.~\ref{fig:selfE}:
\begin{eqnarray}
\Sigma_{i}^{j} (p^{0}, \vec{p}) \equiv \delta_i^j \Sigma(p^0,\vec{p})=-\delta_{i}^{j}\frac{h^{2}}{6\pi \mr}
k^{2} \left(ik+3\mu\right) \ . 
\end{eqnarray}
In this equation, $\mr\equiv \mn\mc/(\mn+\mc)$ is the reduced
mass of the system, $p^{0},\ \vec{p}$ are its total energy and momentum,
and $k^{2}/2\mr\equiv p^{0}-\vec{p}^{2}/{2\mnc}+i\epsilon$ is the
$n$-$c$ system's energy in the C.M. frame.   Here
we use power-divergence subtraction \cite{Ka98A,Ka98B}
to renormalize the loop contribution with $\mu$ as the renormalization
scale. Based on this, we can calculate the fully dressed dimer
propagator $D_{i}^{j}\equiv D\delta_{i}^{j}$ and get the following
\begin{eqnarray}
D^{-1}(p_0,{\bf p})=p_0 - \frac{{\bf p}^2}{2 M_{nc}} +\Delta-\Sigma(p_0,{\bf p})  \ . \label{eqn:propdressing} 
\end{eqnarray}

The general (off-shell) T-matrix can be calculated by using the interaction listed in Eq.~(\ref{eqn:simpleL}), $\mathcal{L}_{P}$, and is then schematically shown as $T\sim V\times D \times V$ with $V$ the $n$-$c$-dimer interaction in the corresponding channel~\cite{Hammer:2011ye}. In the $n$-$c$ C.M. frame we have:
\begin{eqnarray}
\langle \vec{p}'|T(E)|\vec{p} \rangle=\frac{h^{2}}{\mr^{2}} \left(\vec{p}'\cdot \vec{p}\right) \ D(E,{\bf 0}). \label{eqn:TandD}
\end{eqnarray}
By comparing the on-shell T-matrix with the effective range expansion (ERE) \cite{Ka98A,Hammer:2011ye}, we obtain the following relationships ($a$ and $r$ are the scattering volume and effective ``range" in $P$-wave scattering):
\begin{eqnarray}
\Delta \frac{6\pi\mr}{h^{2}}=\frac{1}{a}\ , \quad 
\frac{3 \pi}{h^{2}}+2\mu =-\frac{1}{2} r \ . 
\end{eqnarray} 

To calculate the ANC, we use the following observation \cite{GoldbergerWatson,Hammer:2011ye}:
\begin{align}
\langle \vec{r}^{'}|\frac{1}{E-H}|\vec{r} \rangle&=\langle\vec{r}^{'} |\frac{1}{E-H_{0}}+\frac{1}{E-H_{0}}T\frac{1}{E-H_{0}}|\vec{r} \rangle\notag \\
&\overset{E\rightarrow-B}{\longrightarrow}C^{2}\times \sum_j  \frac{\phi_j (\vec{r}^{'})\phi_j^*(\vec{r})}{E+B} \ . \label{eqn:ancandT}
\end{align}
Here $\phi_j(\vec{r})$ is the zero-range
wave function for a $P$-wave state whose eigenvalue of $J_z$ is $j$. Since spin is absent here:
\begin{equation}
\phi_{j}({\bf r})=\left(1 + \frac{1}{\gamma r}\right) Y_{1j}({\hat r}) \frac{e^{-\gamma r}}{r},
\end{equation}
where $\gamma=\sqrt{2 \mr B}$ is the binding momentum of the bound state, and $C$ is the ANC.

By analytically continuing the calculated T-matrix to negative scattering energy, we find
\begin{equation}
C=\sqrt{\frac{(-)2\gma^{2}}{r+3\gma}} \ .
\end{equation}
As a result, the EFT parameters, $\Delta$ and $h$---or equivalently the scattering parameters $a$ and
$r$---can be fixed using $\gma$ and the ANC, $C$. In principle {\it ab initio} input can be used for both these quantities. However, experimental observables in 
weakly bound systems are extremely sensitive to the two-body binding energy, so in what follows we take $\gma$ from data, and obtain the requisite
ANCs
from an {\it ab initio} method---we use VMC but the same connection can be made to any underlying theory of the $n$-$c$ bound state\footnote{This sensitivity to the separation energy also affects the definitions
of the ANCs themselves.  The ANCs were computed using experimental 
separation energies as inputs, even where these differ from those of the \textit{ab initio}
Hamiltonian \cite{Nollett:2011qf,Nollett:2012}. }.
ANCs can also be measured, in principle,
in transfer-reaction experiments like
those of Ref.~\cite{Trache}. We will discuss the results obtained when experimental input is used for $C$ further below.  We note that similar connections between ANCs and effective-range parameters have previously been obtained without the use of EFT, e.g. in deriving $S$-factor parameterizations of radiative-capture cross sections \cite{YarmukhBaye,BayeBrainis}.

\section{$\lie$ and $\lie^{*}$, and $\lis$-$n$ ($\lis^{*}$-$n$) $P$-wave scatterings} \label{sec:Li8bs}
\begin{figure}
\centering
\includegraphics[scale=0.3, angle=0]{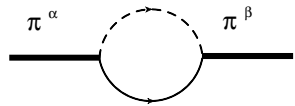}
\caption{The diagram of dimers' self-energy insertions. Here thick and thin solid lines, and the dashed line denote dimers, $n$, and $c$ fields. The dimers include a spinless $P$-wave bound state (Sec.~\ref{sec:example}), $\fd{\pi}{\alpha}$, and $\fd{\tilde{\pi}}{i}$ (Sec.~\ref{sec:Li8bs}).} \label{fig:selfE}
\end{figure}

This section focuses on computing the properties of 
$\lie$ and $\lie^{*}$ as (shallow) bound states within the coupled system consisting of $\lis$-$n$ and $\lis^{*}$-$n$ channels.  These channels have different spin 
structure and mix non-perturbatively. The Lagrangian that describes this is written as:
\begin{eqnarray}
&&\mathcal{L}=\mathcal{L}_0 + \mathcal{L}_S + \mathcal{L}_{P,gs} + \mathcal{L}_{P,es} \ ,\label{eqn:L0}\\
&&\mathcal{L}_{0}=\fdu{n}{\sigma} \left(i\partial_{t}+\frac{\bigtriangledown^{2}}{2\mn} \right) \fd{n}{\sigma} + \fdu{c}{a}\left(i\partial_{t}+\frac{\bigtriangledown^{2}}{2\mc} \right) \fd{c}{a}\notag \\
&& \quad +  \fdu{d}{\delta}\left(i\partial_{t}+\frac{\bigtriangledown^{2}}{2\mc} \right) \fd{d}{\delta} +\fdu{\pi}{\alpha}\left(i\partial_{t}+\frac{\bigtriangledown^{2}}{2\mnc} + \Delta \right) \fd{\pi}{\alpha} \notag \\
&& \quad + \fdu{\tilde{\pi}}{i}\left(i\partial_{t}+\frac{\bigtriangledown^{2}}{2\mnc} + \tilde{\Delta} \right) \fd{\tilde{\pi}}{i} \ , \\
&&\mathcal{L}_{S}= \gone \fdu{c}{a'} \fdu{n}{\sigma'} \T{a'\sigma'}{i} \T{i}{a \sigma} \fd{c}{a}\fd{n}{\sigma} \notag \\
&& \quad +\gtwo \fdu{c}{a'} \fdu{n}{\sigma'} \T{a'\sigma'}{\alpha} \T{\alpha}{a \sigma} \fd{c}{a}\fd{n}{\sigma} \notag \\ 
&& \quad +\gthree \fdu{d}{\delta}\fdu{n}{\sigma'} \T{\delta \sigma'}{i} \T{i}{a\sigma} \fd{c}{a} \fd{n}{\sigma} + \mathrm{C.C.} \ , \label{eqn:L1}\\ 
&&\mathcal{L}_{P,gs}=\hone \fdu{\pi}{\alpha}\T{\alpha}{ij}\T{i}{\sigma a} \fd{n}{\sigma} i\left(\V{n}-\V{c}\right)_{j} \fd{c}{a}\notag \\
&&\quad+\htwo \fdu{\pi}{\alpha}\T{\alpha}{\beta j}\T{\beta}{\sigma a} \fd{n}{\sigma} i\left(\V{n}-\V{c}\right)_{j} \fd{c}{a} \notag \\
&& \quad +\hthree \fdu{\pi}{\alpha} \T{\alpha}{jk} \T{k}{\delta \sigma}\fd{n}{\sigma} i(\V{n}-\V{c^{*}})_{j} \fd{d}{\delta} 
+\mathrm{C.C.} \ , \\
&& \mathcal{L}_{P,es}= \honet \fdu{\tilde{\pi}}{k}\T{k}{ij}\T{i}{\sigma a} \fd{n}{\sigma} i\left(\V{n}-\V{c}\right)_{j} \fd{c}{a}\notag \\
&& \quad +\htwot \fdu{\tilde{\pi}}{k}\T{k}{\beta j}\T{\beta}{\sigma a} \fd{n}{\sigma} i\left(\V{n}-\V{c}\right)_{j} \fd{c}{a} \notag \\
&& \quad +\hthreet \fdu{\tilde{\pi}}{k}\T{k}{0 j}\T{0}{\sigma \delta} \fd{n}{\sigma} i\left(\V{n}-\V{c^{*}}\right)_{j} \fd{d}{\delta} \notag \\
&& \quad +\hfourt \fdu{\tilde{\pi}}{k}\T{k}{i j}\T{i}{\sigma \delta} \fd{n}{\sigma} i\left(\V{n}-\V{c^{*}}\right)_{j} \fd{d}{\delta} +\mathrm{C.C.} \ . \label{eqn:L3}
\end{eqnarray} 
Here fields $n_\sigma$, $\fd{c}{a}$, $\fd{d}{\delta}$, $\fd{\pi}{\alpha}$, and $\fd{\tilde{\pi}}{i}$ correspond to neutron, $\lis$, $\lis^{*}$, $\lie$ and $\lie^{*}$ respectively. The fields' indices are their spin projections with a specific convention: $\sigma, \delta,.. =\pm 1/2$, $a,b,..=\pm 3/2, \, \pm 1/2$, $\alpha,\beta,..=\pm 2, \, \pm 1, \, 0$, and $i,j,..=\pm 1, \, 0$. The $\T{...}{...}$s are Clebsch-Gordan (C-G) coefficients, for example, $\T{\alpha}{ij}$ is the C-G coefficient combining two spin ones to get a spin two, and $\T{ij}{\alpha}$ is, in principle, its complex conjugate, although C-G coefficients are chosen to be real here. The notation for $n$, $\lis$, and $\lis^*$ to couple to the two different $P$-wave dimers is, we hope, obvious. 

We do not introduce $S$-wave dimers. Instead, we reproduce the unnaturally large scattering length in the ${}^5$S$_2$ channel, $a_{({}^5S_2)}=-3.63(5)$ fm~\cite{Koester}, by taking the corresponding $S$-wave coupling $g_{({}^5S_2)}$ to be unnaturally enhanced, i.e. to scale as $1/(M_n \gamma)$~\cite{vK99,Ka98A,Ka98B,Ge98,Bi99}. In contrast, the natural scaling $1/(M_n \Lambda)$ would lead to a scattering length similar to that observed in the ${}^3$S$_1$ channel of ${}^7$Li-$n$ scattering, $a_{({}^3S_1)}=0.87(7)$ fm~\cite{Koester}. (See Table~\ref{table-scattparam} for a list of $S$-wave scattering lengths and  $P$-wave effective ranges, and  their scaling assignments.)  

\vskip -0.2 cm
  \begin{table}[ht!]
  \begin{tabular}{|c|c|c|c|}
 \hline
Parameter & Channel & Value & Assigned scaling\\
 \hline
$a_{({}^5S_2)}$ & $S$-wave, $S=2$ & $-3.63$(5) fm & $1/\gamma$\\

$a_{({}^3S_1)}$ & $S$-wave, $S=1$ & 0.87(7) fm & $1/\Lambda$\\

$r$ & $P$-wave, $J=2$ & $-1.43$(2) fm$^{-1}$ & $\Lambda$ \\

$\tilde{r}$ & $P$-wave, $J=1$ & $-1.86$(6) fm$^{-1}$ & $\Lambda$\\
 \hline
\end{tabular}
\caption{Scattering parameters for $n$-$\lis$ scattering in different channels. The effective ``range" parameters are extracted from VMC calculations, as described in the text around expression~(\ref{eqn:effectiveranges}).}
\label{table-scattparam}
\end{table}

An immediate consequence of the existence of several different channels in the $P$-wave parts of the Lagrangian is that different channels mix, e.g.,  $n+\lis\leftrightarrow n+\lis^{*}$ can occur through the $2^{+}$ and $1^{+}$ channels corresponding to $\lie$ and $\lie^{*}$. As a result, the self-energy insertions of $\fd{\pi}{}$ and $\fd{\tilde{\pi}}{}$ fields, as shown in Fig.~\ref{fig:selfE}, all have contributions from both $n+\lis$ and $n+\lis^{*}$ intermediate states. 

Let's first focus on the $2^{+}$ channel. The self-energy of the $\fd{\pi}{}$ field in this theory is 
\begin{eqnarray}
\Sigma_{\alpha}^{\beta}=\frac{(-)\delta_{\alpha}^{\beta}}{6\pi \mr}\left[\hpt^{2} k^{2} \left(ik+3\mu\right) +\hthree^{2} k'^{2}\left(ik'+3\mu\right) \right]. \nonumber\\
    \label{eqn:selfEpi}
\end{eqnarray}
Here  $k^{2}/2\mr\equiv E $ is the $n$-$c$ energy in the C.M. frame,  $\hpt^{2}\equiv \hone^{2}+\htwo^{2}$, $k'\equiv \sqrt{k^{2}-\gma_{\Delta}^{2}+i\epsilon}$, and $\gma_{\Delta} \equiv \sqrt{2\mr E^{*}}$ with $E^{*}$ the excitation energy of $\lis^\ast$. We also define $\gamma^*$ as the binding momentum of $\lie$ against breakup to $\lis^*$ and a  neutron, so we have $\gma^{*2}=\gma^{2}+\gma_{\Delta}^{2}$. With $E^*=0.478$ MeV and $B=1.05$ MeV we find $\gma=57.8$ MeV, $\gma^*=65.1$ MeV, and $\gamma_\Delta=30.0$ MeV (c.f. Table~\ref{tab:dynamicenergyscales}).

Next, we calculate the $\fd{\pi}{\alpha}$ field's dressed propagator, labeled as $D_{\alpha}^{\beta}\equiv \delta_{\alpha}^{\beta} D$, by summing up all the self-energy insertion diagrams. The result is similar to that in Eq.~(\ref{eqn:propdressing}). However the difference in the consequent T-matrix calculation is that $n+\lis$ now has an inelastic channel for $E > E^*$, i.e. above the threshold for core excitation. Again by using  $T\sim V\times D \times V$, we can compute the general off-shell T-matrix for elastic and inelastic $n+\lis$ and $n+\lis^{*}$ scatterings.  To make a connection to the phase shift analysis in $\lis+n$ elastic scattering channels, we only show the  ($2^{+}$ channel) T-matrix traced over initial- and final-state spins, schematically:
\begin{widetext}
\begin{eqnarray}
\frac{\frac{10 \pi}{\mr} \vec{p}\cdot\vec{p}'}{ \mathrm{Tr}_{\sigma,a} 
\langle \vec{p}';\fd{n}{\sigma},\fd{c}{a}|T(E)|\vec{p};\fd{n}{\sigma},\fd{c}{a}\rangle}= D^{-1}\frac{6\pi\mr}{\hpt^{2}}
=\frac{1}{a} -\frac{\hthree^{2}}{\hpt^{2}} \gma_{\Delta}^{3} - \frac{1}{2}\left(r-3\frac{\hthree^{2}}{\hpt^{2}}\gma_{\Delta}\right)k^{2}
+ i\left[k^{3}+\frac{\hthree^{2}}{\hpt^{2}}(k^{2}-\gma_{\Delta}^{2})^{\frac{3}{2}}\right]. \nonumber\\ \label{eqn:TandD2}
\end{eqnarray}
In Eq.~(\ref{eqn:TandD2})
\begin{align}
-\frac{1}{2} r  \equiv \frac{3\pi}{\hpt^{2}}+(1+\frac{\hthree^{2}}{\hpt^{2}})3\mu-\frac{3\hthree^{2}}{2\hpt^{2}} \gma_{\Delta}; \qquad 
\frac{1}{a} \equiv  \frac{6\pi \mr}{\hpt^{2}}\Delta +\frac{\hthree^{2}}{\hpt^{2}}\gma_{\Delta}^{2}(\gma_{\Delta}-3\mu) \ .
\end{align}
\end{widetext}

At this point we have started from the EFT Lagrangian (\ref{eqn:L0})--(\ref{eqn:L3}) and derived an off-shell $T(E)$ which encodes a modified version of the effective-range expansion appropriate for the
$n$-$\lis$-$\lis^{*}$ Hilbert space; it accounts for the presence of the $\lis^*$-$n$ threshold at momentum $\gamma_\Delta$. 
Several comments are in order here. First, we see that if $k \ll \gma_\Delta$ then the non-analytic factor in Eq.~(\ref{eqn:TandD2}) can be expanded in powers of $k^2$ and we recover the usual ERE in $n$-$\lis$ $2^{+}$ elastic scattering channel:
\begin{equation}
k^{3}\cot{\delta}=-\frac{1}{a}+\frac{1}{2} r k^{2}+ \ldots.  
\end{equation}
This is a manifestation of the decoupling theorem: if additional degrees of freedom in the EFT are at high energies, they may  be replaced by a string of contact operators. But, second, such a treatment prevails only well below the $\lis^{*}$ production threshold. 
Once  $k\geq \gma_{\Delta}$, the inelastic channel is open, with the consequence being an increase in the imaginary part of $D^{-1}$. But, even below the threshold, the non-analyticity in $k^2 - \gma_{\Delta}^2$ will be manifest as rapid dependence of the real part of $D^{-1}$ on $k$. Third, the polynomial dependence for $k \sim \gma_\Delta$ involves coefficients that encode short-distance physics in our EFT.
In particular, the formula for $1/a$ now exhibits a linear dependence on $\mu$, a divergence that comes from the diagram in which $\lie$ is excited to a $\lis^*$-$n$ state. Both this, and the more standard $\lis$-$n$ loop, have a  cubic divergence, but this does not appear in PDS. This linear divergence appears in $1/a$ once the loop contains the additional energy scale, $\gamma_\Delta$.  
For $k \neq 0$, both loops have a linear divergence
that is proportional to $k^2$. The coefficient of the $\mu$-dependent piece of $r$ thus involves a quadratic sum of the couplings to all states to which $\lie$ can couple. Fourth, as a consequence of this, the ERE parameters in our theory have a  different interpretation than those in Refs.~\cite{Rupakprl,Fernando:2011ts}. In this work the specific values of $1/a$ and $r$ emerge as a combination of short-distance ($\sim 1/\Lambda$ effects) and effects at scale $\gamma^*$. 

Equation (\ref{eqn:TandD2}) can be used to analytically continue the $T$ matrix in different channels to negative energy, in order to find the binding energy of $\lie$, from the prescription $D^{-1}(k=i\gma)=0$. By computing the residue of $D$ at this pole we find 
 the wave function renormalization factor ($\sqrt{Z}$) for the $\fu{\pi}{}$ fields up to NLO \cite{Hammer:2011ye}:
\begin{eqnarray}
Z=\frac{(-)6\pi}{\hpt^{2}(r+3\gma)+3\hthree^{2}(\gma^{*}-\gma_{\Delta})} \ .
\label{eqn:renormalization}
\end{eqnarray}
From here we proceed in the same manner as that by which we obtained Eq.~(\ref{eqn:ancandT}) to get the ANCs for $n+\lis$ and $n+\lis^{*}$ in $\lie$: 
\begin{eqnarray}
\frac{C_{(\P{3}{2})}^{2}}{\hone^{2}\gma^{2}}&=&
\frac{C_{(\P{5}{2})}^{2}}{\htwo^{2}\gma^{2}}=
\frac{C_{(\P{3}{2}^{*})}^{2}}{\hthree^{2}\gma^{*2}} =\frac{Z}{3\pi} \ . \label{eqn:Cs}
\end{eqnarray}

The same calculations can be done for the field $\fd{\tilde{\pi}}{}$ that represents $\lie^{*}$. The couplings involved---$\honet$, $\htwot$, $\hthreet$ and $\hfourt$---are listed in the EFT Lagrangian, expression~(\ref{eqn:L3}). The following simple correspondences can be invoked to get formulae for the $1^{+}$ ($\lie^{*}$) channel  which are the analogs of Eqs.~(\ref{eqn:selfEpi})--(\ref{eqn:Cs}) for the $2^{+}$ channel. For couplings: $\Delta\leftrightarrow \tilde{\Delta}$, $h_{(..)}[h_{(..*)}]\leftrightarrow \tilde{h}_{(..)}[\tilde{h}_{(..*)}]$, and for observables: $a\leftrightarrow \tilde{a}$, and $r\leftrightarrow \tilde{r}$. Moreover, in this excited state 
$\sqrt{2 m_R B_{\lie^*}}=41.6$ MeV, while the binding momentum with respect to the $\lis^*$-$n$ threshold is $51.3$ MeV. Once again we have 
$\tilde{\gma}^{*2}\equiv \tilde{\gma}^{2}+\gma_{\Delta}^{2}$ (c.f. Table~\ref{tab:dynamicenergyscales}). 

\begin{widetext}
We denote the dressed propagator for the  $\fd{\tilde{\pi}}{}$ field by $\tilde{D}_{i}^{j}\equiv \delta_{i}^{j} \tilde{D}$, and then get the $T$-matrix (traced over neutron and $\lis$ spins) in the $1^{+}$ channel:
\begin{eqnarray}
\frac{\frac{6 \pi}{\mr} \vec{p}\cdot\vec{p}'}{ \mathrm{Tr}_{\sigma,a} 
\langle \vec{p}';\fd{n}{\sigma},\fd{c}{a}|T(E)|\vec{p};\fd{n}{\sigma},\fd{c}{a}\rangle}= \tilde{D}^{-1}\frac{6\pi\mr}{\htott^{2}}=\frac{1}{\tilde{a}} -\frac{\htotst^{2}}{\htott^{2}} \gma_{\Delta}^{3}  &-&\frac{1}{2}\left(\tilde{r}-3\frac{\htotst^{2}}{\htott^{2}}\gma_{\Delta}\right)k^{2}
+ i\left[k^{3}+\frac{\htotst^{2}}{\htott^{2}}(k^{2}-\gma_{\Delta}^{2})^{\frac{3}{2}}\right]\,\label{eqn:ERE1+}\\
-\frac{1}{2} \tilde{r} \equiv \frac{3\pi}{\htott^{2}}+(1+\frac{\htotst^{2}}{\htott^{2}})3\mu-\frac{3\htotst^{2}}{2\htott^{2}} \gma_{\Delta} \ ;&&
\frac{1}{\tilde{a}}\equiv  \frac{6\pi \mr}{\htott^{2}}\tilde{\Delta} +\frac{\htotst^{2}}{\htott^{2}}\gma_{\Delta}^{2}(\gma_{\Delta}-3\mu) \ ,
\end{eqnarray}
with $\htott^{2}\equiv\honet^{2}+\htwot^{2}$, $\htotst^{2}\equiv \hthreet^{2}+\hfourt^{2}$.  

Again, the conventional ERE for the $1^{+}$ $n$+$\lis$ elastic scattering  channel, $-k^{3}\cot{\tilde{\delta}}=\frac{1}{\tilde{a}}-\frac{1}{2} \tilde{r}k^{2}+.$, holds when $k\ll \gma_{\Delta}$, but Eq.~(\ref{eqn:ERE1+}) accounts for the opening of the ${}^7$Li$^*$-$n$ channel above threshold. In consequence, the $\fd{\tilde{\pi}}{}$ wave function renormalization factor ($\sqrt{\tilde{Z}}$) and the ANCs for $\lie^{*}$ ($\tilde{C}_{(x)}$) at NLO are: 
\begin{eqnarray}
\tilde{Z}&=&\frac{(-)6\pi}{\htott^{2}(\tilde{r}+3\tilde{\gma})+3\htotst^{2}(\tilde{\gma}^{*}-\gma_{\Delta})} \ , \label{eqn:renormalization2} \\
\frac{\tilde{C}_{(\P{3}{1})}^{2}}{\honet^{2}\tilde{\gma}^{2}}=
\frac{\tilde{C}_{(\P{5}{1})}^{2}}{\htwot^{2}\tilde{\gma}^{2}}&=&
\frac{\tilde{C}_{(\P{3}{1}^{*})}^{2}}{\hfourt^{2}\tilde{\gma}^{*2}}=
\frac{\tilde{C}_{(\P{1}{1}^{*})}^{2}}{\hthreet^{2}\tilde{\gma}^{*2}}
=\frac{\tilde{Z}}{3\pi} \ . \label{eqn:tildeCs}
\end{eqnarray}

Equations (\ref{eqn:Cs}) and (\ref{eqn:tildeCs}) indicate we need seven ANCs as input to our LO EFT calculation. In Table~\ref{tab:QMCANCs}, we collect all the computed and measured ANCs  used in the current study.
\begin{table}
   \centering
   \begin{tabular}{|c|c|c|c|c|c|c|c|} \hline
           & $C_{(\P{3}{2})}$
           & $C_{(\P{5}{2})}$   
           & $C_{(\P{3}{2}^{*})}$ 
           & $\tilde{C}_{(\P{3}{1})}$ 
           & $\tilde{C}_{(\P{5}{1})}$
           & $\tilde{C}_{(\P{1}{1}^{*})}$
           & $\tilde{C}_{(\P{3}{1}^{*})}$\\  \hline
      Nollett 
           & $-0.283(12)$
           & $-0.591(12)$ 
           & $-0.384(6)$
           & $0.220(6)$ 
           & $0.197(5)$
           & $-0.195(3)$ 
           & $-0.214(3)$ \\  \hline
 Ref.~\cite{Trache}
           & $ -0.284(23)$
           & $-0.593(23)$ 
           & $$
           & $0.187(16)$ 
           & $0.217(13)$
           & $$ 
           & $$ \\  \hline  
   \end{tabular}
   \caption{ANCs ($\mathrm{fm}^{-\frac{1}{2}}$) for different channels. In the ``Nollett'' ANCs, the $\lis+n$ ANCs can be found Ref.~\cite{Nollett:2011qf}, while $\lis^{\ast}+n$ ANCs are computed by the same methods in this study. The measured $\lis+n$ ANCs are from Ref.~\cite{Trache}.} \label{tab:QMCANCs}
\end{table}
\end{widetext}

On the theory side, the computed ANCs involving the \lis\ ground state were reported in
Ref.~\cite{Nollett:2011qf}. Those involving the excited state were
computed by the same methods and are presented here for the first
time. The ANCs were extracted from wave functions computed by the VMC
method \cite{Wiringa:2009}.    Each wave function was
computed using a Hamiltonian comprising Argonne $v_{18}$ two-nucleon
terms \cite{Wiringa:1995} and Urbana IX three-nucleon terms
\cite{Pudliner:1995}.  While not as precise as Green's function Monte
Carlo (GFMC) wave functions, VMC wave functions are quite accurate for many
purposes (and used as starting points for GFMC).  Accurate calculation
of ANCs from these wave functions is described in
Refs.~\cite{Nollett:2011qf,Nollett:2012}, by a method
briefly recapitulated here.

Direct calculation of an ANC in a many-body model from the definition
given in Eq.~(\ref{eqn:ancandT}) would amount to calculation of an
overlap integral.  Because the variational wave functions are much
harder to optimize in their outer regions than in their interiors,
calculations of ANCs based on overlap integrals are inaccurate
and ambiguous.  This problem is avoided by expressing the ANC as an
integral over the wave function interior, resembling Lippman-Schwinger
calculations of the $T$-matrix.  The ANCs in Table~\ref{tab:QMCANCs}
typically have an error of $<5\%$ due to Monte Carlo sampling, much
less than the capture calculation error estimated below from omission of NLO terms.  There
is a possibly larger but unknown error from the accuracy of the wave
functions and underlying Hamiltonian.  Comparison of the computed ANCs
(and closely-related resonance widths) of many states with
experimental results suggests that this error is typically no
larger than the experimental errors \cite{Nollett:2011qf,Nollett:2012}.
Limited testing with alternatively constructed wave functions supports
this conclusion.

On the experimental side, only $\lis+n$ ANCs have been measured in Ref.~\cite{Trache} \footnote{The original data were presented as squared ANCs in the ``jj''  basis. Here we convert them (and the associated errors) to ``ls'' basis by assuming that all ANCs have the same signs as in the VMC calculations.}. Recently a new measurement has been carried out \cite{Howell2013}, which, within its larger error bars, agrees with that of Ref.~\cite{Trache}. Hence in the following, when using experimental ANCs, we will only mention those from Ref.~\cite{Trache}.

By using the theory ANCs (first line of Table~\ref{tab:QMCANCs}) and Eqs.~(\ref{eqn:renormalization}), (\ref{eqn:Cs}), (\ref{eqn:renormalization2}) and (\ref{eqn:tildeCs}), the effective ranges for the two channels (as well as the $h$ and $\tilde{h}$ couplings) are found. We obtain 
\begin{equation}
r=-1.43(2)~\mathrm{fm}^{-1}; \qquad \tilde{r}=-1.86(6)~\mathrm{fm}^{-1}, \label{eqn:effectiveranges}
\end{equation}
as already quoted in Table~\ref{table-scattparam}.
 Similar effective ranges are found using the measured $\lis+n$ ANCs together with the $\lis^{\ast}+n$ ANCs found via VMC. The magnitude of $r$ and $\tilde{r}$ is larger than the naive high-energy scale $\Lambda \sim 90$ MeV, as already mentioned in Sec.~\ref{sec:intro}.

In this study, we perform a LO calculation of capture observables. Hence we omit terms suppressed by $(\gamma,\gamma^*,\tilde{\gamma})/(r,\tilde{r})$. 
The $\gamma$, $\gma^{\ast}$, 
$\tilde{\gma}$, $\tilde{\gma}^{\ast}$, and $\gma_{\Delta}$ appearing in the denominators of Eqs.~(\ref{eqn:renormalization}) and~(\ref{eqn:renormalization2})
are such effects, so in our LO calculation we drop them, and obtain:
\begin{equation}
Z^{\mathrm{LO}}=\frac{(-)6\pi}{\hpt^{2}r}; \qquad
\tilde{Z}^{\mathrm{LO}}=\frac{(-)6\pi}{\htott^{2}\tilde{r}} \ .
\end{equation}
The LO ANCs, $C^{\mathrm{LO}}_{(\cdots)}$ and $\tilde{C}^{\mathrm{LO}}_{(\cdots)}$, as used in the capture calculations, are then defined through Eqs.~(\ref{eqn:Cs}) and (\ref{eqn:tildeCs}) with $Z,\tilde{Z}\rightarrow Z^{\mathrm{LO}}, \tilde{Z}^{\mathrm{LO}}$.
\begin{eqnarray}
\frac{{C^{\rm LO}_{(\P{3}{2})}}^{2}}{\hone^{2}\gma^{2}}&=&
\frac{{C^{\rm LO}_{(\P{5}{2})}}^{2}}{\htwo^{2}\gma^{2}}=
\frac{{C^{\rm LO}_{(\P{3}{2}^{*})}}^{2}}{\hthree^{2}\gma^{*2}} =\frac{Z^{\rm LO}}{3\pi} \ . \label{eqn:LOCs}\\
\frac{\tilde{C}^{\rm LO}_{(\P{3}{1})}{}^{2}}{\honet^{2}\tilde{\gma}^{2}}&=&
\frac{\tilde{C}^{\rm LO}_{(\P{5}{1})}{}^{2}}{\htwot^{2}\tilde{\gma}^{2}}=
\frac{\tilde{C}^{\rm LO}_{(\P{3}{1}^{*})}{}^{2}}{\hfourt^{2}\tilde{\gma}^{*2}}=
\frac{\tilde{C}^{\rm LO}_{(\P{1}{1}^{*})}{}^{2}}{\hthreet^{2}\tilde{\gma}^{*2}}=
\frac{\tilde{Z}^{\rm LO}}{3\pi} \ .\nonumber\\ \label{eqn:LOtildeCs}
\end{eqnarray}

Given EFT couplings, ANCs determined from Eqs.~(\ref{eqn:LOCs}) and (\ref{eqn:LOtildeCs}) are smaller than those from Eqs.~(\ref{eqn:Cs}) and (\ref{eqn:tildeCs}) by an amount that is formally of order $\gamma/r \sim 1/5$, but in actuality is somewhat larger, due to the factor ``3'' in front of $\gamma$ and $\tilde{\gamma}$ in expressions~(\ref{eqn:renormalization}) and~(\ref{eqn:renormalization2}). 

\section{Radiative neutron capture} \label{sec:capture}
\begin{figure}
\centering
\includegraphics[scale=0.35, angle=0]{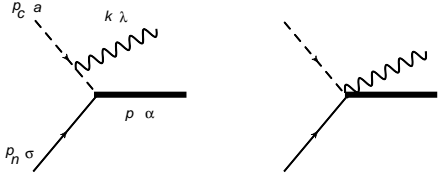}
\caption{Tree diagrams for neutron-capture to $\lie$ and $\lie^{*}$. The line assignments for different fields are explained by the momentum and spin labellings. These diagrams are at LO for both initial total spin $S_i=2$ and $S_i=1$ channels. The dominant components in the initial state are $\S{5}{2}$ and $\S{3}{1}$, but $D$ wave components also contribute in the left diagram.} \label{fig:ncapture_tree}
\end{figure}

\begin{figure}
\centering
\includegraphics[scale=0.45, angle=0]{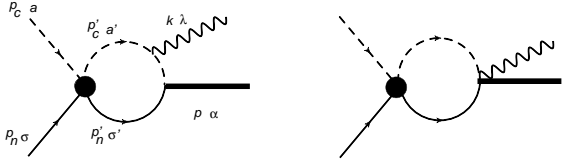}
\caption{Loop diagrams for neutron-capture to $\lie$ and $\lie^{*}$. The black blob corresponds to the scattering of the incoming particles in the $S$-wave, as shown in Fig.~\ref{fig:swavescattering}. The sum of the two diagrams is finite~\cite{Hammer:2011ye, Rupakprl}. Only the $S_{i}=2$ channel contributes at LO, while perturbative initial-state scattering effects in the $S_{i}=1$  channel enter at NLO. With the $\fd{c}{}$ fields in the loop changed to $\fd{d}{}$ fields these diagrams represent the dominant dynamical effect of core excitation, which occurs at NLO in the $\gamma/\Lambda$ expansion. Detailed discussions can be found in the text.} \label{fig:ncapture_loop}
\end{figure}

\vskip -0.2 cm
\begin{figure}
\centering
\includegraphics[scale=0.5, angle=0]{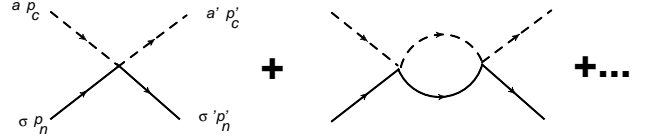}
\caption{The diagrams for the multiple $S$-wave scatterings. They are shown as a black blob in Fig.~\ref{fig:ncapture_loop}.} \label{fig:swavescattering}
\end{figure}

In this section, we present our LO calculation of capture to $\lie$ and $\lie^{*}$. A general incoming state can be decomposed to states of  initial spin $S_i=2$ and $S_i=1$. To calculate the total cross section, we add the partial cross sections from these different incoming states {\it incoherently}. 

We first study capture to $\lie$. The tree-level diagrams at LO are shown in Fig.~\ref{fig:ncapture_tree}. There are both $S$- and $D$-wave components in $S_i=2$ and $S_i=1$ channels.  In the loop diagrams shown in Fig.~\ref{fig:ncapture_loop} only the incoming $S$-wave contributes: $D$-wave initial-state interactions are suppressed by five orders in the EFT expansion. The $S_{i}=1$ loop diagrams also do not appear until NLO because the incoming wave scattering length in that channel  $\aone=0.87(7) \ \mathrm{fm}$ \cite{Koester} is natural, corresponding to a coupling $\gone$ which scales with $1/\Lambda$. Therefore the only loop diagrams needed at LO are for $n$-$\lis$ $S_{i}=2$  channel, where there is an unnaturally large incoming wave scattering length $\atwo=-3.63(5) \ \mathrm{fm}$ \cite{Koester}. Diagrams like Fig.~\ref{fig:ncapture_loop}, but with core excitation in the intermediate state are only allowed for the $\lis$-$n$ $S_{i}=1$ channel, and so are also at least NLO. They will be discussed further below.

\begin{widetext}
For the $\lis + n(S_i=2)\rightarrow \lie(2^{+})+\gma$ partial
cross section the capture amplitude $\mathcal{M}$ at LO including
the diagrams in Fig.~\ref{fig:ncapture_tree}
and (non-core-excitation diagrams of) Fig.~\ref{fig:ncapture_loop} is
\begin{eqnarray}
&& \mathcal{M}=ie_{c} \htwo \sqrt{8Z^{\mathrm{LO}}\mn\mc\mnc}\T{\beta}{\sigma a} \T{\alpha}{\beta j}  \left[\frac{\es(\lambda)\cdot\V{c}}{p_{c}^{0}-\omega-\frac{\left(\pc-\vec{k}\right)^{2}}{2\mc}+i\epsilon} \left(\frac{\pc}{\mr}-\frac{\vec{k}}{\mc}\right)_{j} +(1+X(p_c;\gamma,\atwo) \frac{\es(\lambda)_{j}}{\mc}\right], \nonumber\\ \label{eq:amplexpr}  \\
&& \sum_{\sigma,a}^{\alpha,\lambda} |\mathcal{M}|^{2}=\frac{5}{3}  64 \pi \alpha Z_{c}^{2} \frac{3\pi}{ \gma^{2}} \frac{\mn^{2}}{\mr}  \left(C_{(\P{5}{2})}^{\mathrm{LO}}\right)^{2} \left[|1+X(p_c;\gamma,\atwo)|^{2}-\frac{2 \pc^{2}\sin^{2}{\theta}}{\pc^{2}+\gma^{2}} \left(\frac{\gma^{2}}{\pc^{2}+\gma^{2}}+ \mbox{Re}\left\{X(p_c;\gamma,\atwo)\right\}\right)\right]. \nonumber\\ \label{eqn:txsec1}
\end{eqnarray} 
In Eq.~(\ref{eq:amplexpr}), $\sigma$, $a$, and $\alpha$ are the spin projections of the neutron, $\lis$, and $\lie$, while $\es(\lambda)$ is the photon field vector with polarization $\lambda$.  The initial core momentum is $\pc$, the photon final momentum is $\vec{k}$ (both in the C.M. frame), and $\theta$ is the angle between them. The factors $\alpha$ and $Z_{c}$ are the electromagnetic fine structure constant and the proton number of the core, so $e_c = Z_c |e|$. The LO wave-function renormalization factors, $Z^{\mathrm LO}$, and ANCs, $C_{(..)}^{\mathrm{LO}}$, were defined in Sec.~\ref{sec:Li8bs}. Meanwhile, $X$ 
is a function that encodes loop contributions for capture from an $S$-wave state of relative momentum $p$, where there is a large scattering length, $a$, to a $P$-wave state with binding momentum $\gamma$. It is defined as:
\begin{equation}
X(p_c;\gamma,a) \equiv  \frac{(-)i}{a^{-1}+ip_{c}} \left[p_{c}-\frac{2}{3}i\frac{\gma^{3}-ip_{c}^{3}}{\gma^{2}+p_{c}^{2}}\right]  \ . \label{eqn:xtwodef}
\end{equation}
For the $\lis + n(S_i=1)\rightarrow \lie(2^{+})+\gma$ partial cross section, we can go through the same calculations and get $|\mathcal{M}|^{2}$ with initial and final quantum numbers ($i,f$) summed up:
\begin{eqnarray}
\sum_{i,f} |\mathcal{M}|^{2}&=&\frac{5}{3}  64 \pi \alpha Z_{c}^{2} \frac{3\pi}{ \gma^{2}} \frac{\mn^{2}}{\mr}  \left(C_{(\P{3}{2})}^{\mathrm{LO}}\right)^{2} \left[1-\frac{\pc^{2}\sin^{2}{\theta}}{\pc^{2}+\gma^{2}} \frac{2\gma^{2}}{\pc^{2}+\gma^{2}} \right] \ , \label{eqn:txsec2}
\end{eqnarray} 
where, because $\aone/\atwo=O(\gamma/\Lambda)$, the loop diagram is NLO in this channel. The factor $5/3$ in both Eqs.~(\ref{eqn:txsec1}) and~(\ref{eqn:txsec2}) comes from summing up all the spin indices in $\T{\beta}{\sigma a} \T{\alpha}{\beta j} \T{\sigma a}{\beta'}\T{\beta' j'}{\alpha}=\T{\alpha}{\beta j}\T{\beta j'}{\alpha}=5/3\delta_{j'}^{j}$ ($S_i=2$ channel), and in $\T{i}{\sigma a} \T{\alpha}{i j} \T{\sigma a}{i'}\T{i'j'}{\alpha}=\T{\alpha}{ij}\T{i j'}{\alpha}=5/3\delta_{j'}^{j}$ ($S_i=1$ channel). 

For capture to $\lie^{*}$, we have parallel calculations and results:
\begin{eqnarray}
\sum_{i,f} |\mathcal{M}|^{2}&=& 64 \pi \alpha Z_{c}^{2} \frac{3\pi}{\gmat^{2}} \frac{\mn^{2}}{\mr}  \bigg\{\left(\tilde{C}_{(\P{3}{1})}^{\mathrm{LO}}\right)^{2} \left[1-\frac{\pc^{2}\sin^{2}{\theta}}{\pc^{2}+\gmat^{2}} \left(\frac{2\gmat^{2}}{\pc^{2}+\gmat^{2}}\right)\right] \notag \\
&& \qquad +\left(\tilde{C}_{(\P{5}{1})}^{\mathrm{LO}}\right)^{2} \left[|1+X(p_c;\gmat,\atwo)|^{2}-\frac{2 \pc^{2}\sin^{2}{\theta}}{\pc^{2}+\gmat^{2}} \left(\frac{\gmat^{2}}{\pc^{2}+\gmat^{2}}+\mbox{Re}\left\{X(p_c;\gmat,\atwo)\right\}\right)\right] \bigg\} \ . \label{eqn:txsec3}
\end{eqnarray}
Here $X$ appears again due to the loop diagrams (Fig.~\ref{fig:ncapture_loop}) for the $S_i=2$ capture to the $\lie$ excited state, and the loop effect in the $S_{i}=1$ channel is NLO and not included here. The factor $\frac{5}{3}$ in this case is absent, because here the spin summation is $\T{i}{\sigma a} \T{k}{i j} \T{\sigma a}{i'}\T{i'j'}{k}=\T{k}{ij}\T{i j'}{k}=\delta_{j'}^{j}$ ($S_i=1$) and $\T{\beta}{\sigma a} \T{k}{\beta j} \T{\sigma a}{\beta '}\T{\beta 'j'}{k}=\T{k}{\beta j}\T{\beta j'}{k}=\delta_{j'}^{j}$ ($S_i=2$).

Now let's consider the effect of core excitation in the diagrams shown in Fig.~\ref{fig:ncapture_loop}. The fact that the total spin is unaffected by the multiple scattering in the incoming wave means that this effect can only occur if the initial $n$-$\lis$ total spin $S_i=1$.  For example, for capture to $\lie$, by doing calculations similar to those presented above, we find that core excitation modifies the loop effect from $X(p_c;\gamma,\aone)$ to $X(p_c;\gamma,\aone) + \frac{\hthree}{\hone}Y$, where 
\begin{eqnarray}
Y &\equiv& \gthree  \frac{\mr}{2\pi} \bigg[-\sqrt{\gma_{\Delta}^{2}-p_{c}^{2}-i\epsilon}+ \frac{2\bigg(\gma^{*3}-(\gma_{\Delta}^{2}-p_{c}^{2})\sqrt{\gma_{\Delta}^{2}-p_{c}^{2}-i\epsilon}\bigg)}{3(\gma^{2}+p_{c}^{2})}\bigg], \nonumber\\
&\overset{p_{c}\rightarrow 0}{\longrightarrow}& \gthree \frac{\mr \gamma}{2\pi} \left[\frac{2}{3} \sqrt{1 + \frac{\gamma_\Delta^2}{\gamma^2}}\left(1 + \frac{\gamma_\Delta^{2}}{\gamma^{2}}\right) - \frac{\gamma_\Delta}{\gamma} - \frac{2}{3} \frac{\gamma_\Delta^3}{\gamma^3}\right],
\end{eqnarray}
with the coupling $\gthree$ defined in Eq.~(\ref{eqn:L1}). The two core-excitation diagrams, when summed together, are convergent: their ultraviolet behavior is the same as the analogous diagrams without core excitation~\cite{Hammer:2011ye}.
\end{widetext} 

The factor $Y$ can change the cross section in the $S_{i}=1$ channel. However, if we assume that $\gthree$ is natural, i.e. not abnormally enhanced (c.f. the $\S{5}{2}$ incoming channel), then $Y \sim \gamma/\Lambda$ as $p_c \rightarrow 0$. This is to be compared with the loop function $X$ which goes to $\gamma a$ as $p_c \rightarrow 0$. This is also $\gamma/\Lambda$ for a natural scattering length (as in the $\S{3}{1}$) but is $\sim 1$ if $a$ is unnaturally large (as in the $\S{5}{2}$).  The naturalness of $\gthree$ is supported by results in Ref.~\cite{knox87,Navratil:2010jn} which show that that the inelastic $\lis(n,n^\prime)\lis^\ast$
cross section is much smaller than that for elastic scattering from threshold up to $E_{n}$ of a few MeV. However, it is possible that in other systems the  inelastic scattering is unnaturally enhanced. Core-excitation loop diagrams should be included in LO calculations in such systems. Even in the absence of dynamical core-excitation effects, such as those in Fig.~\ref{fig:ncapture_loop}, the possibility to excite the ${}^7$Li core to ${}^7$Li$^*$ affects our LO calculation: it alters the distribution of strength between different channels in the dimer self-energy, thus changing the relationship between, e.g. ANCs and the effective range, $r$, in the $\lie$ channel. 

We can address the decoupling theorem for the core-excitation diagrams by taking the (unphysical) limit, $\gma/\gma_{\Delta}\sim p_{c}/\gma_{\Delta} \rightarrow 0$, in which case we get
\begin{eqnarray}
Y\longrightarrow \gthree \frac{\mr}{2\pi} O(\frac{p_{c}^{2},\gma^{2}}{\gma_{\Delta}})\,. 
\end{eqnarray}
This indicates that if the energy scale of core excitation is high, its dynamical impact at low energy is suppressed to N$^2$LO, which is in accord with the power counting for 
the E1 contact operator associated with an incoming $S$-wave channel with a natural scattering length~\cite{Hammer:2011ye}.

In contrast, there is a contact term associated with the E1 $\S{5}{2} \rightarrow \lie$ transition already at NLO. We can see this by the following simple argument.
Because this is an E1 transition, the ratio between the contact contribution and the full result in the $\S{5}{2}$ channel can be estimated as \cite{Hammer:2011ye}
\begin{eqnarray}
\frac{\int_{0}^{1/\Lambda} dr (r-a)r(1+\frac{1}{\gma r}) e^{-\gma r}}{\int_{0}^{+\infty} dr (r-a)r(1+\frac{1}{\gma r}) e^{-\gma r}} \sim O\left(\frac{\gamma}{\Lambda}\right).
\end{eqnarray}

\begin{figure}
\centering
\includegraphics[scale=0.6, angle=-90]{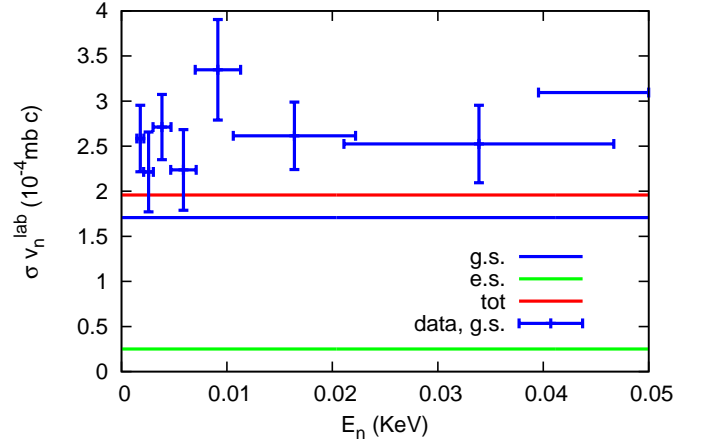}
\caption{Total cross section $\times$ neutron velocity vs.~neutron lab energy. ``g.s.'' and  ``e.s.'' correspond to capture to $\lie$ and $\lie^{*}$, while ``tot'' is the sum of these two. The calculation is based on the computed ANCs. The data are from Ref.~\cite{Blackmon}.} \label{fig:xsecvel}
\end{figure}

\begin{figure}
\centering
\includegraphics[scale=0.6, angle=-90]{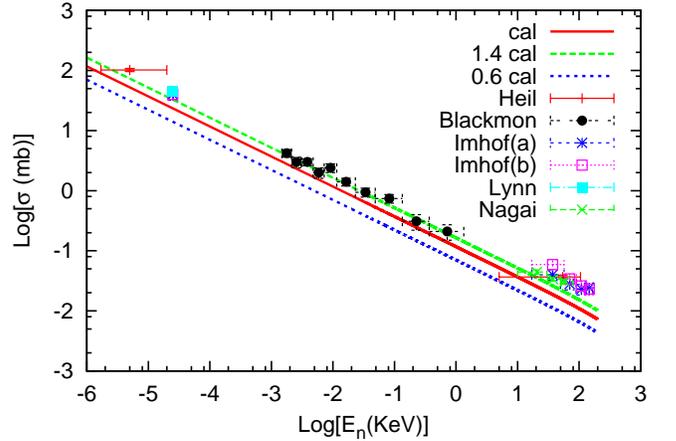}
\caption{Total cross section based on computed ANCs vs.~neutron lab energy. 
The data are from Refs.~\cite{Blackmon,Barker,Lynn,Nagai,knox87,Heil,Imhof}. The ``Imhof (a)'' and ``Imhof (b)'' are the same data from Ref.~\cite{Imhof} normalized to two different reference cross sections.  Three different calculation curves are shown. ``cal'' is our LO results, while ``1.4cal'' and ``0.6cal'' are the LO results multiplied by 1.4 and 0.6 respectively, so as to indicate the uncertainty of the LO Halo EFT prediction for the cross section.} \label{fig:xsec}
\end{figure}

Now we have all the ingredients needed to calculate the total cross section at LO, using the LO ANCs and the following formula:
\begin{eqnarray}
d\sigma = \frac{1}{64 \pi^{2} \mnc^{2}} \frac{\omega}{p_{c}} \frac{1}{8}\sum|\mathcal{M}|^{2} d\Omega
\end{eqnarray}
We collect our results in Figs.~\ref{fig:xsecvel} and~\ref{fig:xsec}. 
We can see that $1/v$ behavior of the calculated cross section holds up to $E_{n}=200$ keV. The nominal accuracy of our LO amplitude is $\sim \gamma/r \approx 20$\%. This translates into an uncertainty of $\approx 40$\% for the cross section. There is also a much smaller uncertainty ($< 5$\%) in the cross-section prediction due to the uncertainties in the VMC ANCs. Using the measured ANCs~\cite{Trache}---where available---instead of VMC ANCs to obtain values for $r$ and $\tilde{r}$ leads to cross sections that differ from those shown in Fig.~\ref{fig:xsecvel} by only a few per cent.
An error band indicating the overall 40\% uncertainty in the LO Halo-EFT calculation is shown in Fig.~\ref{fig:xsec}. We find agreement between theory and experiment within the combined error bars. The central value for the threshold $\sigma v$ that is predicted by the theory is below the mean value of the data, however we already see that corrections $\sim \gamma/r$ to the ANCs used in the EFT calculation will push the theory prediction higher.


In addition to total cross sections, the ratio of partial cross sections associated with 
different initial spin states and branching ratios to different final states
have been measured.  
Our LO results based on the computed ANCs agree very well with these data on relative amplitudes.
Considering first the relative contributions of different initial spin states, $p_c=0$
gives
\begin{eqnarray}
\frac{\sigma[(S_i=1)\rightarrow 2^{+}]}{\sigma[(S_i=2)\rightarrow 2^{+}]}=  \frac{\left(C_{(\P{3}{2})}^{\mathrm{LO}}\right)^{2}}{\left(C_{(\P{5}{2})}^{\mathrm{LO}}\right)^{2}(1-\frac{2}{3}\gma \atwo)^{2}} \ . \notag 
\end{eqnarray}  
Therefore
\begin{eqnarray}
\frac{\sigma[(S_i=2)\rightarrow 2^{+}]}{\sigma(\rightarrow 2^{+})} = 0.927 , 
\label{eq:spin2tototalgroundstate}
\end{eqnarray}
where $\sigma(\rightarrow J^\pi)$ denotes the total cross section to the 
specified final state including all initial spin states. This---as well as the expressions for other threshold ratios given below---hold as long as $p_c \ll \gamma, \tilde{\gamma}$. 

The ratio (\ref{eq:spin2tototalgroundstate}) is largely unaffected by higher-order effects in the ANCs: the ratio $C_{(\P{3}{2})}/C_{(\P{5}{2})}$ is fixed by the relative size of the couplings $h$ to these two channels and does not change when NLO terms are included in the denominator of (\ref{eqn:renormalization}). Once dynamical core excitation appears at NLO it will affect the rate for capture from the $\S{3}{1}$, but not that from the $\S{5}{2}$. Initial-state interactions also appear at NLO in the $S=1$ channel. We can use the size of this second effect to estimate the impact of NLO corrections on the result (\ref{eq:spin2tototalgroundstate}). Including $X(0;\gamma,a_{\S{3}{1}})$ in the expression for the cross-section ratio alters the number given in (\ref{eq:spin2tototalgroundstate}) by 2\%, so this prediction appears to be  very stable against higher-order effects. Meanwhile, the use of experimental (rather than VMC) ANCs as input to determine $r$ produces a consistent result within the experimental uncertainties on the ANCs. We thus obtain a final result, including higher-order uncertainties:
\begin{eqnarray}
\frac{\sigma[(S_i=2)\rightarrow 2^{+}]}{\sigma(\rightarrow 2^{+})} = 0.93(2).
\label{eq:spin2tototalgroundstatefinal}
\end{eqnarray}
In Ref.~\cite{Gulko} (c.f.~Ref.~\cite{Barker}), a lower bound of 0.86 for the 
ratio $\sigma[{(S_i=2)}\rightarrow 2^{+}]/\sigma(\rightarrow 2^{+})$ has 
been reported. Our result is certainly consistent with that constraint. In contrast, the EFT 
calculation in Ref.~\cite{Rupakprl} 
assumed equal
$n$-$\lis$ coupling strengths in $\P{3}{2}$ and $\P{5}{2}$ channels
and consequently failed
to satisfy this experimental lower bound. 

There is no experimental data on the corresponding ratio of $S_{i}=2$ and $S_{i}=1$ partial cross sections for capture to the excited state. We predict this ratio to be 
\begin{equation}
\frac{\sigma[(S_i=2)\rightarrow 1^{+}]}{\sigma(\rightarrow 1^{+})} = 0.65
\label{eq:spin2tototalexcitedstate}
\end{equation}
using the VMC ANCs as input. The uncertainty in the input affects that prediction by $< 1$\%. This is the only case in which the use of experimental ANCs alters the result significantly: we obtain 0.75(4) for the ratio if we use the ANCs of Ref.~\cite{Trache} as input, with the error bar stemming from the sizable uncertainties in the ANC measurements of that work. The difference with Eq.~(\ref{eq:spin2tototalexcitedstate}) is driven by the different values of the  ANCs $\tilde{C}_{(\P{3}{1})}$ and $\tilde{C}_{(\P{5}{1})}$. Meanwhile, arguing in the same manner by which we obtained the uncertainty quoted in Eq.~(\ref{eq:spin2tototalgroundstatefinal}), we estimate higher-order effects in the excited-state $S_i=2$ and $S_i=1$ partial cross section ratio to be about 9\%. We emphasize that this uncertainty is separate from the larger uncertainty stemming from discordant values for the input to the EFT calculation. Therefore here we quote two numbers:
\begin{equation}
\frac{\sigma[(S_i=2)\rightarrow 1^{+}]}{\sigma(\rightarrow 1^{+})} = 0.65(6) \; \mbox{ or } \; 0.75(7),
\label{eq:spin2tototalexcitedstatefinal}
\end{equation}
where the first is a prediction using VMC ANCs as input and the second uses the ANCs measured in Ref.~\cite{Trache} to set the size of $\tilde{r}$. Both results include both the uncertainty due to higher-order Halo-EFT effects and uncertainty due to the input used in our LO EFT calculation. We note that a measurement of this ratio  at the higher end of the range shown in Eq.~(\ref{eq:spin2tototalexcitedstatefinal}) would speak in favor of the larger ANCs quoted in Ref.~\cite{Trache}, while, conversely, a spin-2-to-total excited-state capture cross-section ratio in the neighborhood of 0.6 would suggest that the VMC ANCs are correct.

Next, we discuss the branching ratio for capture to the ground state.
Near threshold, the ratio of capture cross sections is
\begin{eqnarray}
\frac{\sigma(\rightarrow 1^+)}{\sigma(\rightarrow 2^+)}&=&\frac{3}{5} \frac{\left(\tilde{C}^{\rm LO}_{(\P{3}{1})}\right)^2 + \left(\tilde{C}^{\rm LO}_{(\P{5}{1})}\right)^2 |1 - \frac{2}{3} \atwo \tilde{\gamma}|^2}{\left(C^{\rm LO}_{(\P{3}{2})}\right)^2 + \left(C^{\rm LO}_{(\P{5}{2})}\right)^2 |1 - \frac{2}{3} \atwo \gamma|^2}    \notag \\
&\Rightarrow& \frac{\sigma(\rightarrow 2^+)}{\sigma}=0.88, \label{eq:1plusto2plus}
\end{eqnarray}
a result which is reflected in Fig.~\ref{fig:xsecvel}. This ratio is largely controlled by the excited-state and ground-state ANCs, but is also affected by the initial state (rescattering) effect due to large $S$-wave scattering length in $S_i=2$ channels [cf. Eqs.~(\ref{eqn:txsec1}), (\ref{eqn:txsec2}), and~(\ref{eqn:txsec3})]. 
In Ref.~\cite{Lynn}, the 
branching 
ratio is measured to be $0.89\pm 0.01$ for thermal neutrons. The authors of 
Ref.~\cite{Nagai}
also found $0.89\pm 0.01$ at 20 to 70 keV. Both of these measurements are in excellent agreement with our number.

To estimate the NLO effect we use the full ANCs, defined by Eq.~(\ref{eqn:Cs}), rather than the LO ones. This changes the result for the branching ratio by about 5\%, although it should be noted that the ratio $\sigma(\rightarrow 1^+)/\sigma(\rightarrow 2^+)$ is reduced by the nominal 40\% of an NLO effect, it's just that there is very little contribution from capture to the $1^+$ to begin with. Effects due to initial-state scattering in the spin-1 channel and dynamical core excitation will also enter at NLO, and are expected to have similar, $\sim 40$\%, impact on $\sigma(\rightarrow 1^+)/\sigma(\rightarrow 2^+)$. These may compensate somewhat for the reduction in the ratio (\ref{eq:1plusto2plus}) due to NLO effects in the ANCs.
Thus, overall we have
\begin{equation}
\frac{\sigma(\rightarrow 2^+)}{\sigma}=0.88(4)
\end{equation}
The error bar due to uncertainties in the EFT input (ANCs) is negligible compared to that from NLO effects.

\section{Summary} \label{sec:sum}

We have studied $\lis(n,\gma)\lie$
in the framework of Halo-EFT, using ANCs from \textit{ab initio} calculations to fix EFT parameters. One could also use ANCs inferred from transfer-reaction experiments for this purpose; we have checked that doing so in the $\lie$ system does not alter any of our conclusions substantially. 
Our total cross section result shows agreement with the available data at the level expected of a calculation at leading order in the $\gamma/\Lambda$ expansion of Halo EFT. 
Interestingly the branching ratio between the threshold captures to $\lie$ and to $\lie^{*}$, as well as the ratio of partial cross sections originating from $S_i=2$ and $S_i=1$ initial spin states agree very well with data. The first ratio is controlled by the 
quadratic sum of  \lie\ and $\lie^\ast$ ANCs, together with initial-state ($\S{5}{2}$) rescattering.
The second ratio is proportional to the squared
ratio of $\P{3}{2}$ and $\P{5}{2}$ ANCs
in $\lie$ and is also modified by the strong initial-state effect. In comparison, the 
previous
EFT calculation \cite{Rupakprl,Fernando:2011ts}, which assumed 
equal
$n$-$\lis$ coupling strengths in $\P{5}{2}$ and $\P{3}{2}$ channels, fails to satisfy the measured lower bound on the fraction of captures that proceed from the $S_i=2$ initial state.
We also calculated the
ratio of partial cross sections for capture to $\lie^*$ with different initial-state spins. No data are presently available for this observable. All of these ratios should have smaller higher-order corrections than will the total cross section.

Another advance reported here is the inclusion of nonperturbative channel mixing via the inclusion of core excitation in the effective-range expansion for the two-body $T$-matrix. The behavior of the effective-range expansion in the presence of coupled channels was also studied in the context of EFT in Ref.~\cite{LenskyBirse}, but there the focus was on the case for scattering with $l=0$.  
The effective-range expansion for the case we studied here, where there are multiple channels with different thresholds, was discussed---albeit in very different formalisms to ours---in Ref.~ \cite{RossShaw,Rakityansky}; Refs.~\cite{Biedenharn,Blokhintsev} considered 
effective-range expansions for multiple open channels with the same threshold.

The general $T$-matrix results obtained here should hold in any system with strong $P$-wave scattering and a low-lying excitation of the core. Dynamical core excitation is accidentally suppressed in the $\lis$-$n$ system, so that it only appears at NLO, but it can be a leading-order effect in other cases.
In particular, we emphasize that the effective-range expansion used in Refs.~\cite{Rupakprl,Fernando:2011ts} has a radius of convergence $p < \gamma_\Delta$, and so formally it does not permit analytic continuation of scattering data to the $\lie$ pole. Thus, although our result $r=-1.43(2)~{\rm fm}^{-1}$ is quite close to that found by fitting the capture cross section in Ref.~\cite{Rupakprl}, $r=-1.47~{\rm fm}^{-1}$, only in our calculation can the connection between the bound-state ANC and the effective-range expansion parameter $r$ be properly established. Furthermore,
our explicit inclusion of $\gamma_\Delta$ as a low-energy scale means that the number we obtain for $r$ has distinct contributions from physics at scale $\Lambda$ and effects due to the $\lis$ excited state. 

The agreement with data is promising and encourages study of higher-order contributions for this process in the same approach. It also suggests that it may be worthwhile to include the $3^+$ resonance explicitly in the EFT, as was done in Ref.~\cite{Fernando:2011ts}, so that we are not limited to examining capture cross section data at $E_n \leq 200$ keV.
However, higher-order calculations will involve more undetermined EFT constants, and additional strategies to fix these from {\it ab initio} calculations and/or data will be required.
The present calculation benefited from the close connection between ANCs, which have been of considerable interest to the nuclear structure community in recent years, and LO couplings in Halo-EFT.  Less obvious correspondences between the two formalisms will have to be found before higher-order parameters can be fixed, or else work will be needed to identify measured observables that can meet the same need. 

This methodology can also be applied to study radiative capture to other shallow two-body bound states. This could be especially useful in cases where data is scarce.
We are presently applying these methods to the reaction $^7\mathrm{Be}(p,\gma)^8\mathrm{B}$. This is closely related to $^7$Li neutron capture by isospin symmetry in microscopic models, but the two processes present rather different challenges for Halo-EFT because of the Coulomb interaction.  We expect the relationship between Halo-EFT calculations of these two isospin-mirror reactions will be informative.

\section*{Acknowledgements}
X.Z. and D.R.P. acknowledge support from the US Department of Energy under grant DE-FG02-93ER-40756. K.M.N. acknowledges support from the Institute of Nuclear and Particle Physics at Ohio University.

\end{document}